# Decision curve analysis for personalized treatment choice between multiple options

Konstantina Chalkou, MSc[1,2], Andrew J. Vickers, PhD[3], Fabio Pellegrini, PhD[4], Andrea Manca, PhD[5], Georgia Salanti, PhD[1]

**Affiliation:** Institute of Social and Preventive Medicine, University of Bern, Bern, Switzerland [1]; Graduate School for Health Sciences, University of Bern, Switzerland [2]; Department of Epidemiology and Biostatistics, Memorial Sloan-Kettering Cancer Center, New York, NY, USA[3]; BDH, Biogen Spain, Madrid, Spain [4]; Centre for Health Economics, University of York, York, UK [5]

*Key words* decision curve analysis, network meta-analysis, prediction model, net benefit, clinical usefulness

## Abstract

**Background:** Decision curve analysis can be used to determine whether a personalized model for treatment benefit would lead to better clinical decisions. Decision curve analysis methods have been described to estimate treatment benefit using data from a single RCT.

**Objectives**: Our main objective is to extend the decision curve analysis methodology to the scenario where several treatment options exist and evidence about their effects comes from a set of trials, synthesized using network meta-analysis (NMA).

**Methods:** We describe the steps needed to estimate the net benefit of a prediction model using evidence from studies synthesized in an NMA. We show how to compare personalized versus one-size-fit-all treatment decision-making strategies, like 'treat none' or 'treat all patients with a specific treatment' strategies. First, threshold values for each included treatment need to be defined (i.e., the minimum risk difference compared to control that renders a treatment worth taking). The net benefit per strategy can then be plotted for a plausible range of threshold values to reveal the most clinically useful strategy. We applied our methodology to an NMA prediction model for relapsing-remitting multiple sclerosis, which can be used to choose between natalizumab, dimethyl fumarate, glatiramer acetate, and placebo.

**Results:** We illustrated the extended decision curve analysis methodology using several threshold values combinations for each available treatment. For the examined threshold values, the 'treat patients according to the prediction model' strategy performs either better than or close to the one-size-fit-all treatment strategies. However, even small differences may be important in clinical decision-making. As the advantage of the personalized model was not consistent across all thresholds, improving the existing model (by including, for example, predictors that will increase discrimination) is needed before advocating its clinical usefulness.

**Conclusions:** This novel extension of decision curve analysis can be applied to NMA based prediction models to evaluate their use to aid treatment decision-making.

*Highlights*

- Extension of the decision curve analysis into a (network) meta-analysis framework

- Evaluation of personalized models predicting treatment benefit when several treatment options are available and evidence about their effects comes from a set of trials.

- Detailed steps to compare personalized versus one-size-fit-all treatment decision-making strategies

- This extension of decision curve analysis can be applied to (network) meta-analysis-based prediction models to evaluate their use to aid treatment decision-making.

# 1 Introduction

Randomized controlled trials (RCTs) and their meta-analyses have traditionally focused on inferences about treatment effects for the average patient. Yet what clinicians want to know is the treatment effect for the patient in front of them, and the effects of treatment may differ between individuals.[1] To identify the best treatment option for an individual, researchers can employ prediction models to evaluate the treatment effects on health outcomes as a function of patient-level characteristics.[2,3,4,5]

Personalized prediction models could be used to identify groups of patients for which the benefits of treatment outweigh the harms. Doing so would require extensive validation and such validation should include an evaluation of clinical utility. The latter refers to the ability of the model to guide treatment decisions at the point of care. While methods to evaluate a model's performance have been well studied and are described in the literature (e.g., calibration measures, the area under the curve (AUC), etc.),[6,7,8] evaluation of the clinical utility of a model is a relatively new concept.

Decision curve analysis (DCA) has been proposed to evaluate the clinical utility of personalized prediction models, and does not aim to replace other analyses with different objectives, such as a cost-effectiveness analysis.[9,10] DCA can be applied to models that predict an absolute risk (such as a model to predict the risk of cancer to guide decisions about biopsy) and those predicting treatment benefit (such as a model to predict the change in outcome associated with drug therapy).[9,11,12] The data used to calculate net benefit for a treatment strategy, typically come from an RCT that compares two treatments: a reference treatment (such as no treatment or placebo) and an active treatment of interest. The DCA methodology, its use, as well as several tutorials can be found in https://www.mskcc.org/departments/epidemiology-biostatistics/biostatistics/decision-curve-analysis.

There are often several treatment options for a given condition. Unfortunately, there is often uncertainty about their relative benefits due to lack of direct head-to-head comparison in a single RCT. Evidence synthesis in the form of pairwise meta-analysis (PMA) and its extension, network meta-analysis (NMA), can be used both to structure the evidence base (summarising direct and indirect comparisons) and to produce an estimate of the effects of any treatment against other available options. It has been found that prediction models based on (network) meta-regression of multiple individual patient data (IPD) can be used to identify the best treatment option for an individual patient.[11,13,14,15]

Consider a patient diagnosed with relapsing-remitting multiple sclerosis (RRMS) who is contemplating starting a disease-modifying drug. The individual and her clinician may have access to the results of an NMA of aggregated data to inform their decision, but this evidence only gives insight into the expected health outcomes and the efficacy of the treatments being considered for the 'average patient' in the model.[16,17,18] Personalized treatment recommendations can be obtained if patient characteristics are taken into account when predicting the outcome under different treatment options. This can be achieved using network meta-regression with IPD data,[13] with the model indicating the optimal drug (in the case of RRMS treatment decision, this may be the one that minimizes the predicted risk to relapse over the time horizon of two years) for any given patient profile. Extending this idea to several outcomes and accounting for the trade-off between safety and efficacy, will result in a hierarchy of treatment options that is tailored to a participant's characteristics.[19,20]

In this paper, we extend the DCA methodology, as proposed by Vickers et. al.,[10] to evaluate the clinical usefulness of a personalized prediction model that aims at recommending a treatment among many possible options according to individual characteristics, such as the one described above. The focus of the paper is methodological, and we use an example with RRMS only to outline the developed methodology. This work is supported and funded by the HTx. The HTx is a Horizon 2020 project supported by the European Union lasting for 5 years from January 2019. The main aim of HTx is to create a framework for the Next Generation Health Technology Assessment (HTA) to support patient-centered, societally oriented, real-time decision-making on access to and reimbursement for health technologies throughout Europe. We describe the network meta-regression prediction model in Section 2. In Section 3 we describe ways to select the threshold values for each treatment option, how treatment recommendations can be made based on the results of a network meta-regression prediction model, and we describe the estimation of quantities in DCA methodology from PMA and NMA datasets. We show the results from the case study in Section 4, and we conclude with a discussion of the advantages and limitations of the proposed approach.

## 2 Case study: Personalized treatment recommendation for patients with relapsing-remitting multiple sclerosis

Multiple sclerosis (MS) is an immune-mediated disease of the central nervous system with several subtypes. The most common subtype is RRMS.[21] Patients with RRMS present with acute or subacute symptoms (relapses) followed by periods of complete or incomplete recovery (remissions).[22] Effective treatment of patients with RRMS can prevent disease progression and associated severe consequences, like spasticity, fatigue, cognitive dysfunction, depression, bladder dysfunction, bowel dysfunction, sexual dysfunction, pain and death.[23] There are several available treatment options for RRMS [16] and their efficacy and safety profiles vary. For instance, natalizumab is more effective (on average) than dimethyl fumarate, but associated with important side-effects and increased risk of progressive multifocal leukoencephalopathy, which can cause death.[24,25]

Recently a two-stage model was presented to predict personalized probability of relapse within two years in patients diagnosed with RRMS.[13] Three phase III RCTs were used: AFFIRM, DEFINE, and CONFIRM.[26,27,28] Patients were randomized into three active drugs (natalizumab, glatiramer acetate, dimethyl fumarate) and placebo as shown in **Figure 1**.

In a *first stage*, the baseline risk score for relapse was developed, that is a score which summarizes the patient-level characteristics and indicates the baseline health-condition's severity. In a *second stage*, the baseline risk score was used as the only effect modifier, which impacts on relative treatment effects, in a network meta-regression model to predict the risk to relapse within the next two years under the three drugs or placebo.[13] The results are presented in **Figure 2**, as well as in an interactive R-Shiny application available at https://cinema.ispm.unibe.ch/shinies/koms/. A detailed description of the development of the RRMS personalized prediction model, which we use as an example here, has been previously given.[13]

Such models can be used to guide clinical decisions, assuming heuristically that relapse is the only health outcome of interest. For example, this prediction model would recommend dimethyl fumarate to patients whose baseline risk is lower than 25% and natalizumab to patients whose baseline risk is higher than 25%. However, even when a patient has baseline risk score equal to 30%, where natalizumab minimizes the predicted risk to relapse, the absolute predicted difference in relapse probability is only 5% compared to dimethyl fumarate. In addition, natalizumab is a drug with more serious side-effects compare to dimethyl fumarate, and hence, the doctor in discussion with the patient might decide for dimethyl fumarate.

We want to evaluate whether this personalized prediction model could guide the decision-making process. We will compare the treatment decisions that this model entails ('treat patients according to the prediction model') to those from 'one-size-fit-all' strategies: 'treat none', 'treat all patients with natalizumab', 'treat all patients with dimethyl fumarate', 'treat all patients with glatiramer acetate'.

## 3 Methods

In section 3.1, we describe how treatment recommendations via a prediction model are reached, when we have multiple treatment options. In section 3.2 we introduce the proposed extension of the DCA methodology when considering several competing treatment strategies. In section 3.3, we describe the implementation and software to evaluate the model on predicting the optimal treatment to prevent relapsing within the next two years in RRMS.

### 3.1 Threshold values

Let us consider there are several treatment options available for a health condition. Each available treatment option is denoted with $j$, where $j = 1, 2, ..., J$. Each treatment is associated with different side-effects, cost and inconvenience. Say treatment A is administred as a pill, has no serious side-effects, has good tolerability and low cost, whereas injectable treatment B is less safe and more expensive. If A and B are equally effective on positive health outcomes, then A would be preferred. If B is more effective than A, then the choice of B over A will depend on how much more effective it is to justify the extra risk, cost and inconvenience. The required effectiveness of B or A so that B is chosen, is what the threshold value measures. For a dichotomous beneficial outcome, the threshold value $T_j$ for treatment $j$ is defined as the minimum risk difference compared to control treatment that renders treatment $j$ worth taking. $T_j$ sums into a single value: the harms, costs and inconvenience of treatment $j$ and expresses how much benefit would be expected to outweigh the harm that treatment $j$ might cause. At a population level, setting $T_j$=20% means that we would be willing to treat up 5 patients with $j$ to prevent one patient relapsing; four patients will be taking unnecessarily the drug (and hence being subjected in its toxicity) and are traded against one patient with prevented relapse. In the RRMS example, it would be reasonable to set a lower threshold for dimethyl fumarate and glatiramer acetate compared to natalizumab, ($T_{DF}$%) because of their different side-effect profiles. As different patients might weight differently the risk of an event and risks associated with each treatment, a clinically relevant range of threshold values for all treatment options is required.[29] To set this range, a patients' survey may be planned. Researchers can elicit individual preferences from a study sample and integrate utilities across the distribution of these preferences in a more formal decision analysis that will lead into a better justification of the threshold range.[29] In the RRMS example, we used a range of threshold values based on discussions with two experienced MS neurologists (see acknowledgements) on drugs' side-effects and toxicity to illustrate how the suggested methodology could be applied.

*3.2   Reaching treatment recommendations when we have multiple options via a model*

Let us consider a personalized prediction model for the probability of an event, $R_{i,j}$, for each patient $i$, where $i$=1, 2, …, N, under each available treatment option $j$, where $j = 1, 2, ..., J$. Then, the risk difference, $RD_{i,j}$, for patient $i$ between treatment $j$ and the control treatment (or placebo) is the difference between the patient's predicted probabilities under these two options: $RD_{i,j}=R_{i,control} - R_{i,j}$. Whether patient $i$ will be prescribed treatment $j$ depends on several factors. First, treatment $j$ needs to be effective, $RD_{i,j} > 0$, that is, it must decrease the predicted probability of a harmful outcome compared to the control treatment.

Second, the benefits of treatment need to outweigh its harms. For example, natalizumab is a drug with important side-effects and associated with increased mortality.[24, 25] Now, imagine a RRMS patient, whose predicted risk to relapse within two years is decreased by $RD_{i,N} = 3\%$ under natalizumab compared to placebo. It is possible that, given the side-effects of treatment, this patient will not choose natalizumab for such a small reduction in predicted probability of relapse.

We define the threshold value $T_j$ as the minimum risk difference compared to control that renders treatment $j$ worth taking. $T_j$ depends on the risks, harms, costs and inconvenience of treatment $j$. For a patient $i$, the recommended treatment $j$ under the prediction model is the one that satisfies $max\{RD_{i,j} - T_j\}$, between those treatments with $RD_{i,j} \geq T_j$. When all active treatments lead to $RD_{i,j} < T_j$, then the control treatment is recommended for patient $i$. $T_{DF} = T_{GA} = 10\%$ and $T_N = 20\%$ In **Table 1**, we present a fictional example showing how treatment recommendation is made via a prognostic model. In this example, the predicted $RD_{i,DL}$ of Dimethyl Fumaratedimethyl fumarate and $RD_{i,N}$ of Natalizumabnatalizumab are larger than their corresponding thresholds, and the maximum difference between the risk difference and the corresponding threshold value is that of Dimethyl Fumarate.dimethyl fumarate.

While the model makes personalized predictions under each treatment $j$, the threshold values $T_j$ are not based on individual preferences.[29] To evaluate the clinical usefulness of the model, we first need to understand the typical range of preferences of patients, with respect to the possible trade-off between harms and benefits of each treatment. Then, these preferences will determine the range of thresholds over which the clinical utility of the model comparing the various competing strategies should be assessed.[29]

### 3.3 *Comparing different treatment strategies, via Decision Curve Analysis*

In the case of medical treatments, there are several decision strategies that can be evaluated and compared. Consider a treatment strategy $s$ that refers to the choice between $j = 0,1, ... J$ treatments, with 0 denoting the control. That strategy recommends a treatment for each patient and can be 'treat all with drug $j$' ($s = j$), 'treat none' ($s = 0$) or a more nuanced strategy suggested by a prediction model. A strategy associated with a prediction model was discussed in detail in the previous section; now assume that the recommended treatment for a patient is well defined in each of the $s$ competing strategies. Control could be any treatment or combination of treatments used as reference, e.g., standards of care, placebo, or no treatment at all. From now and on we will assume placebo as control and strategy $s = 0$ as 'treat none' strategy.

The measure of performance of each strategy is *the net benefit* (NB). NB is the benefit that a decision entails minus the relevant harms weighted by a trade-off preference value. In the case of medical treatments, *benefit* could be measured as the reduction in a harmful health outcome (e.g., relapses) with the treatment. *Harms* include all disbenefits of treatment, including side-effects, risks, financial cost and inconvenience. Vickers et al. described in detail the DCA methodology and defined the net treatment benefit for a single treatment.[10] The NB estimation involves counterfactuals, the unobserved outcome if a particular strategy is employed. Consequently, estimation of NB for a model predicting treatment benefit is best estimated using RCT data.[10]

We generalize the idea to the net benefit of a strategy $s$, $NB_s$, for two or more treatment options, and we show how to estimate it in a PMA and NMA of RCTs. We define $NB_s$ as the benefit (decrease in event rate using strategy $s$) minus the treatment rates multiplied by a set of treatment-specific threshold values $T_j$. The threshold values $T_j$ are measured on a risk scale (from 0 to 1) and identifies which reduction in risk will justify the use of each treatment. Notice that the value of $T_j$ may vary from patient to patient depending on personal preferences and other medical considerations (such as comorbidities). The strategy $s$ with the highest net benefit, for specific threshold values $T_j$, is chosen as leading to better clinical decisions.[10]

More specifically, we define the *net benefit* of each strategy $s$ compared to strategy $s = 0$ ('treat none') as:

$$NB_s = \varepsilon_0 - \varepsilon_s - \sum_j \pi_{s,j} \times T_j$$

where $\varepsilon_0$ denotes the event rate under no treatment, $\varepsilon_s$ the event rate under strategy $s$, and $\pi_{s,j}$ the proportion of patients treated with treatment $j$ under strategy $s$.

### 3.3.1 Estimation of $\varepsilon_0$

When data from *one RCT* with placebo are available, $\varepsilon_0$ is directly quantifiable from the data as the observed proportion of participants with an event in the placebo arm $\hat{\varepsilon}_0 = e_0^{Data}$, where *Data* is the dataset of all available RCTs.[10] However, when we have *several RCTs instead of one*, estimation needs to account for the fact that patients are randomized within trials but not across them. Hence, when estimating event rates we cannot simply pool treatment arms together or results will be biased (Simpson's Paradox).[30,31] In this case, we first need to perform a meta-analysis of all placebo event rates in *Data* across trials, to obtain an estimate of the pooled event rate in the placebo $\hat{\varepsilon}_0$.

### 3.3.2 Estimation of $\pi_{s,j}$ and $\varepsilon_s$

The interest now lies in the estimation of $\varepsilon_s$ and $\pi_{s,j}$ with strategy $s$ when several RCTs are available that compare different subsets of the treatments. This is accomplished by considering the congruent dataset for strategy $s$, $Data_s$. A congruent dataset for $s$, is the subset of *Data* including those patients for whom the recommended treatment coincides with the actual given treatment. Using $Data_s$, we estimate all $\pi_{s,j}$ as the observed proportion of participants under each treatment $j$ in strategy $s$, $\hat{\pi}_{s,j} = p_{s,j}^{Data_s}$.

To derive $\varepsilon_s$ we need the following steps. First, we need to estimate the event rate $\varepsilon_{s,j}$ for each treatment as recommended by strategy $s$. Then, the weighted average event rate under strategy $s$, can be estimated as

$$\hat{\varepsilon}_s = \sum_{j=0}^{J} p_{s,j}^{Data_s} \times \hat{\varepsilon}_{s,j}$$

The quantity $\hat{\varepsilon}_{s,j}$ depends on the strategy and the available data.

1. When we have *only one RCT,* then $\hat{\varepsilon}_{s,j} = e_j^{Data_s}$ i.e., $\varepsilon_{s,j}$ is estimated as the observed proportion of events under arm $j$ in $Data_s$.

2. When we have several RCTs, we first need to perform a meta-analysis of all placebo arms in $Data_s$ to obtain an estimate of the pooled placebo event rate $\hat{\varepsilon}_{s,0}$. Then we perform a synthesis of all studies in $Data_s$ to estimate the pooled risk ratio of each treatment versus the control, $RR_j^{Data_s}$. Then, the treatment-specific event rates are $\hat{\varepsilon}_{s,j} = \hat{\varepsilon}_{s,0} \times RR_j^{Data_s}$.

  a. In case, $Data_s$ does not include placebo arms (e.g., when the treatments are highly effective or when the threshold values set are very low is more likely for the model to recommend an active treatment rather than placebo, and therefore the congruent dataset will include only active treatment arms), we could estimate the pooled event rate $\hat{\varepsilon}_{s,k}$ for another treatment $k$ (instead of $\hat{\varepsilon}_{s,0}$), designated as the reference treatment, included in the congruent dataset. Then we again perform a synthesis of all studies in $Data_s$ to estimate the pooled risk ratio of each treatment versus $k$ treatment, $RR_j^{Data_s}$. Then, the treatment-specific event rates are $\hat{\varepsilon}_{s,j} = \hat{\varepsilon}_{s,k} \times RR_j^{Data_s}$

  b. When $Data_s$ includes only one treatment arm, we could estimate the event rate $\hat{\varepsilon}_{s,j}$ as a meta-analysis of all $j$ arms in $Data_s$.

3. When the strategy $s$ is *'treat all with treatment $j$=x'*, with x≠0, the event rate $\hat{\varepsilon}_{s,x}$ can be estimated from the entire dataset $Data$ as $\hat{\varepsilon}_{s,x} = \hat{\varepsilon}_x = \hat{\varepsilon}_0 \times RR_x^{Data}$. The observed proportion $p_{s,x}^{Data_s}$ is equal to 1, whereas the observed proportion $p_{s,j\neq x}^{Data_s}$ is equal to 0.

4. When the strategy $s$ is *'treat none'*, then the NB is 0 as $\hat{\varepsilon}_{s,0} = \hat{\varepsilon}_0$, and the $\hat{\pi}_{s,j} = p_{s,j}^{Data_s}$ is 0 for all the available treatments $j$.

Hence, considering the nature of the strategies, the congruent dataset, $Data_s$, is mainly used when the NB of a personalized model needs to be estimated. For the NB estimation of all other "fit all" strategies the entire dataset, $Data$, is used. Considering also the nature of the available data, when several RCTs comparing several treatments are available, NMA and/or meta-analysis need to be performed for the NB estimation; however, when only one RCT is available, the observed proportion of the event can be directly estimated.

### 3.3.3 Net Benefit and comparisons of strategies

We define the net benefit, which can be applied to all strategies and settings (i.e., one RCT, several RCTs, single treatment comparison, and several treatment comparisons) as:

$$NB_s = \varepsilon_0 - \sum_{j=0}^{J} \pi_{s,j} \times \varepsilon_{s,j} - \sum_{j=0}^{J} \pi_{s,j} \times T_j$$

The $NB_s$ ranges between $-\max\{T_j\}$ and 1. It is $-\max\{T_j\}$ when there is no decrease in event rate compared to 'treat none' and at the same time all patients take the drug with the highest threshold value $T_j$. Net benefit has a theoretical maximum of 1 for the impossible case where the decrease in event rate is 100% and none of the patients takes any treatment.

The advantage of any strategy $s = w$ compared to a strategy $s = m$, for specific threshold values $T_w$ and $T_m$, can be calculated as the difference between the $NB_w$ and the $NB_m$, and can be interpreted in terms of decrease in event rate as: the use of strategy $w$ compared to strategy $m$ leads to $NB_w - NB_m$ fewer events for a constant treatment rate in each treatment $j$.

All the required steps to calculate the NB for several strategies into an NMA of the RRMS example are in detail presented in Table 2. Following these steps, the NB for each strategy $s$ is estimated for each combination of threshold values. If the personalized model has the highest NB across the entire range of threshold values, then its clinical relevance compared to the default strategies can be argued. If the optimal approach depends on the threshold values, then the typical conclusion would be that the personalized model is of unproven benefit. [29]

### 3.4 *Application in comparison of treatment strategies in RRMS*

We used net benefit to evaluate the clinical usefulness of the two-stage personalized prediction model (described briefly in section 2). As natalizumab is a treatment with serious side-effects and less safe than the other two options, patients would be prescribed natalizumab only when their benefit (here the predicted risk difference) is high. Dimethyl fumarate and glatiramer acetate on the other hand are similar in terms of side-effects and considered safer than natalizumab. In line with two consulting MS neurolostis, the threshold for natalizumab was set higher than those for dimethyl fumarate and glatiramer acetate.

We first assume a threshold value $T_N$=20% for natalizumab and an equal (and lower) threshold value for the other two treatments, $T_{GA}=T_{DF}$=10%. These threshold values reflect the drugs' profiles. Different patients might weight in differently the risk to relapse, and the risks associated with each treatment. Therefore, we consider a range of threshold values for natalizumab (19% - 40%) in combination with a range of common threshold values for dimethyl fumarate and glatiramer acetate (4% - 25%). These ranges were selected based on safety concerns for each drug and on the congruent dataset's limitations; for lower threshold values, the NB could not be calculated because the congruent dataset had only single arm studies, and hence NMA could not be conducted. Then we plot $NB_s$ as a function of threshold values $T_j$, to identify which treatment strategy leads to better clinical decision under different preferences on threshold values.

All our analyses were done in R,[32] using R 3.6.2 version. We make the code available in a GitHub library: https://github.com/htx-r/Reproduce-results-from-papers/tree/master/DCA_NMA. The analysis code uses the `metaprop` command to estimate the event rate in control arm and the `netmeta` command to estimate the relative risk for each active treatment versus placebo.

## 4  Results

The results from comparing the five competing strategies with treatment thresholds $T_N = 20\%$, $T_{DF} = T_{GA} = 10\%$ are presented in **Table 3**, following all the steps presented in **Table 2**. A more detailed description for these estimations is presented in the **Appendix**. Using these thresholds, the strategy based on the prediction model will recommend dimethyl fumarate to 1251 patients, natalizumab to 740 patients, and placebo to 9 patients. No patient is recommended to take glatiramer acetate. For the example threshold values ($T_N = 20\%$, $T_{DF} = T_{GA} = 10\%$), the congruent dataset with the model strategy includes 652 patients; 4 patients in placebo, 418 in dimethyl fumarate, and 230 in natalizumab. The $NB_s$ values are presented in **Table 2** and show that treating RRMS patients using strategy 'treat patients according to the prediction model' results to $NB_s$ higher compared to the default "one-fit-all" strategies. The $NB_s$ for the 'treat patients according to the prediction model' strategy is equal to 0.17.

In **Figure 3**, we present the $NB_s$ for each strategy when $19\% \leq T_N \leq 40\%$ and $T_{GA}=T_{DF}=10\%$. This is an introductory plot, which connects the traditional way of presenting the DCA results with its suggested extension, and shows that the strategy 'treat patients according to the prediction model' has the highest $NB_s$ compared to the other strategies, almost in the whole range of natalizumab threshold values. We were restricted to use as minimum threshold natalizumab value this of 19% as lower threshold values result to a congruent dataset that consists only of single-arm studies, and hence NMA cannot be conducted. For threshold values higher than 40% the results remain the same and 'treat based on the model' strategy outperforms the others.

In Figure 4, we also present a heat-plot showing the strategy with the highest $NB_s$ when $T_N$ is between 19% and 40% in combination with $T_{GA}=T_{DF}$ ranging between 4% and 25%. The empty grey cells in Figure 4 correspond to $T_{GA}=T_{DF} > T_N$ that is deemed clinically irrational. The numbers in the cells are differences in NB between the two strategies (multiplied with 100). As our focus is the clinical utility of the personalized prediction model Figure 4 presents the NB from the model versus the highest NB from the default strategies. For instance, when $T_{GA}=T_{DF} = 20\%$ and $T_N$ 25%, 'treat all with dimethyl fumarate' strategy outperforms all other strategies with a NB difference (multiplied with 100) versus 'treat patients according to the prediction model' strategy of 0.2. This means that treating everyone with dimethyl fumarate would lead to 0.2% fewer relapse events compared to if we chose treatment based on the model. The strategy 'treat patients according to the prediction model' performs either better than or close to the one-size-fit-all treatment strategies (based on the NB differences). However, even small differences may be important in clinical decision-making. The strategy 'treat patients according to the prediction model' lead to better clinical decisions, when the thresholds for dimethyl fumarate and glatiramer acetate are low (<~10%), or when the threshold value for natalizumab is low (<~22%). 'Treat none' strategy seems to outperform the others, when all threshold values are high (i.e., for natalizumab >25%, for dimethyl fumarate and glatiramer acetate >20%). 'Treat all with dimethyl fumarate' strategy, seems to lead to better clinical decisions, when the thresholds for dimethyl fumarate and glatiramer acetate are intermediate (between 10% and 20%), and at the same time the threshold for natalizumab is high (>25%). The strategy 'treat all patients with glatiramer acetate' does not lead to the largest NB for any of the examined threshold combinations. Our methodology raises some questions about the universal applicability of the current personalized model and indicates that a better personalized model may be needed to be universally applicable for decision-making.

## 5 Discussion

We extended the DCA methodology to an NMA framework to evaluate the clinical usefulness of a prediction model that aims at recommending a treatment among many possible options according to individual characteristics[10,33] The personalized prediction models are used to inform patients and decision-makers about the most appropriate treatment for each patient and hence contribute to personalised medicine[13,14,34] Such models need to be evaluated for their ability to guide treatment decisions at the point of care. For this purpose, Vickers et. al. proposed DCA, which is the tool to evaluate such prediction models by comparing the benefit-risk trade-offs they entail to those of other default treatment strategies or other available personalized prediction models.[10] The data used to evaluate such prediction models typically come from an RCT that compares two treatments: a reference treatment and the treatment of interest. As the treatment options for each condition are numerous and their effects are evaluated in multiple RCTs, the extended proposed DCA approach could contribute to evaluate the ability of the widely used personalized prediction models to guide treatment decisions. We applied our methodology for RRMS to evaluate the strategy of choosing between three disease modifying drugs (natalizumab, dimethyl fumarate, glatiramer acetate) and placebo using a personalized prediction model.[13]

The methods and their application in the dataset of treatments for RRMS have several limitations. The personalized prediction model compares only three active drugs among all available options (more than 15 available). The same approach can be applied to personalized prediction models that compare all relevant competing drugs, assuming that studies that compare them are available. The main limitation of our application is the inefficient dataset's sample size; estimation of the parameters needed to estimate the net benefit in our approach needs a large amount of data to ensure that the sample size of the congruent dataset will be large enough to conduct NMA. Confidence intervals around the estimated NB could be shown to present uncertainty due to the limited sample size,[35] however they are not typically used within a classical decision-making approach.[29]

Another technical issue is that it is possible that the congruent dataset for some thresholds includes many single-arm studies. In our application, we omitted the single arm studies from the NMA in the congruent dataset in order to establish causal effects of the treatments, but this resulted in discarding potentially relevant information. When the congruent dataset consists of single arms, (network) meta-analysis cannot be conducted at all. Consequently, NBs cannot be estimated for some threshold combinations, and researchers have to calculate lower and upper bounds for the thresholds examined to ensure that they would lead to enough data to estimate NB. In our application, the lower bounds were outside the range of thresholds indicated by the expert neurologists as relevant. In practice however, the lack of suitable data to estimate NB for relevant thresholds can limit the applicability of DCA. The issue of single-arm studies in the congruent dataset should be subject of further research. Models that include single arm studies in the meta-analysis could be considered, although the risk of bias in the estimates they provide is not to be underestimated.[36,37,38,39,40] Finally, the strategies need to be evaluated for a relevant range of threshold values for all treatment options as different patients might weight differently the risk of an event and risks associated with each treatment.[29] In our application, we defined equal threshold values for dimethyl fumarate and glatiramer acetate and higher threshold values for natalizumab based on expert opinion of two MS neurologists based on the drugs' safety profiles. In practical application, integration of utilities across a distribution of patients' preferences might be used to justify the relevant threshold values range.[1]

To our knowledge, this is the first attempt to evaluate the clinical usefulness of a prediction model that refers to multiple treatments and, consequently, uses evidence from several studies that compare subsets of the competing treatments, relying on the assumptions undelying NMA and prediction models (transitivity, consistency, correct model specification etc.).[41,42,43,44] The proposed approach can be used to compare several treatment strategies and we show how to estimate the net benefit of a treatment strategy using causal treatment effects. If the strategy based on the personalized model is shown to be clinically useful compared to the default 'treat all patients with X' strategy, this does not necessairily mean that it should be implemented in practice. In many clinical areas, the treating physician is evaluating the patient and decides about the treatment strategy without a guiding tool. This state-of-art strategy, needs to be compared to the strageby based on the model in a clinical trial, to inform about the health benefits, patients's experiences and costs associated with clinical implementation.

Personalized prediction models for treatment recommendation have recently gained ground and their popularity will increase with the availability of more data. It is therefore important that such models are evaluated for their performance before they are ready for use by decision-makers. The traditional biostatistical metrics of calibration and discrimination can be useful for analysts to determine how to build and evaluate a model, but cannot determine its clinical value.[8,10,29] We have contributed to the existing methodological arsenal by providing a method to infer about a prediction model's clinical utility in a (network) meta-analysis framework. With the proposed approach, and assuming that enough data from several randomized trials would be available, the evaluation of clinical relevance will now be possible for several prediction models comparing many treatment options.

**Acknowledgements:** See Title Page for Acknowledgements

**Funding statement:** See Title Page for Funding Statement

**Conflicts of interest:** See Title Page for Conflicts of interest

**Data availability statement:** The data that support the findings of this study were made available from Biogen International GmbH. Restrictions apply to the availability of these data, which were used under license for this study.

## 6  References

1.  Kent DM, Paulus JK, van Klaveren D, et al. The Predictive Approaches to Treatment effect Heterogeneity (PATH) Statement. *Ann Intern Med* 2020; 172: 35–45.

2.  Rekkas A, Paulus JK, Raman G, et al. Predictive approaches to heterogeneous treatment effects: a scoping review. *BMC Med Res Methodol* 2020; 20: 264.

3.  Rothwell PM. Can overall results of clinical trials be applied to all patients? *Lancet Lond Engl* 1995; 345: 1616–1619.

4.  Kent DM, Hayward RA. Limitations of Applying Summary Results of Clinical Trials to Individual Patients: The Need for Risk Stratification. *JAMA* 2007; 298: 1209–1212.

5.  Varadhan R, Segal JB, Boyd CM, et al. A framework for the analysis of heterogeneity of treatment effect in patient-centered outcomes research. *J Clin Epidemiol* 2013; 66: 818–825.

6.  The meaning and use of the area under a receiver operating characteristic (ROC) curve. | Radiology, https://pubs.rsna.org/doi/abs/10.1148/radiology.143.1.7063747 (accessed 17 March 2021).

7.  Evaluating the Yield of Medical Tests | JAMA | JAMA Network, https://jamanetwork.com/journals/jama/article-abstract/372568?casa_token=28zbFFHW8v0AAAAA:rP4a0ZrV6hUc3khcgGWcSNoSZVnuCYGUVTPC34oOCyKq0TImtt0iNu3Tb4hT9YLKVEvGip0rp2J6SQ (accessed 17 March 2021).

8.  Steyerberg EW, Vickers AJ, Cook NR, et al. Assessing the performance of prediction models: a framework for traditional and novel measures. *Epidemiol Camb Mass* 2010; 21: 128–138.

9.  Vickers AJ, Elkin EB. Decision Curve Analysis: A Novel Method for Evaluating Prediction Models. *Med Decis Making* 2006; 26: 565–574.

10. Vickers AJ, Kattan MW, Sargent DJ. Method for evaluating prediction models that apply the results of randomized trials to individual patients. *Trials* 2007; 8: 14.

11. Kent DM, Steyerberg E, van Klaveren D. Personalized evidence based medicine: predictive approaches to heterogeneous treatment effects. *BMJ* 2018; k4245.

12. Baker T, Gerdin M. The clinical usefulness of prognostic prediction models in critical illness. Eur J Intern Med 2017; 45: 37–40.

13. Chalkou K, Steyerberg E, Egger M, et al. A two-stage prediction model for heterogeneous effects of treatments. Stat Med 2021; 40: 4362–4375.

14. Seo M, White IR, Furukawa TA, et al. Comparing methods for estimating patient-specific treatment effects in individual patient data meta-analysis. *Stat Med* 2021; 40: 1553–1573.

15. Belias M, Rovers MM, Reitsma JB, et al. Statistical approaches to identify subgroups in meta-analysis of individual participant data: a simulation study. *BMC Med Res Methodol* 2019; 19: 183.

16. Tramacere I DGC, Salanti G DR, Filippini G. Immunomodulators and immunosuppressants for relapsing-remitting multiple sclerosis: a network meta-analysis. *Cochrane Database Syst Rev 2015*; CD011381. DOI: 10.1002/14651858.CD011381.pub2.

17. Lucchetta RC, Tonin FS, Borba HHL, et al. Disease-Modifying Therapies for Relapsing-Remitting Multiple Sclerosis: A Network Meta-Analysis. *CNS Drugs* 2018; 32: 813–826.

18. Fogarty E, Schmitz S, Tubridy N, et al. Comparative efficacy of disease-modifying therapies for patients with relapsing remitting multiple sclerosis: Systematic review and network meta-analysis. *Mult Scler Relat Disord* 2016; 9: 23–30.

19. Naci H, Fleurence R. Using indirect evidence to determine the comparative effectiveness of prescription drugs: Do benefits outweight risks? *HealthOutcResMed* 2011; 2: 241–249.

20. van Valkenhoef G, Tervonen T, Zhao J, et al. Multicriteria benefit-risk assessment using network meta-analysis. *JClinEpidemiol* 2012; 65: 394–403.

21. Ghasemi N, Razavi S, Nikzad E. Multiple Sclerosis: Pathogenesis, Symptoms, Diagnoses and Cell-Based Therapy. *Cell J Yakhteh* 2017; 19: 1–10.

22. Goldenberg MM. Multiple Sclerosis Review. *Pharm Ther* 2012; 37: 175–184.

23. Crayton HJ, Rossman HS. Managing the symptoms of multiple sclerosis: A multimodal approach. *Clin Ther* 2006; 28: 445–460.

24. Rafiee Zadeh A, Askari M, Azadani NN, et al. Mechanism and adverse effects of multiple sclerosis drugs: a review article. Part 1. *Int J Physiol Pathophysiol Pharmacol* 2019; 11: 95–104.

25. Hoepner R, Faissner S, Salmen A, et al. Efficacy and Side Effects of Natalizumab Therapy in Patients with Multiple Sclerosis. *J Cent Nerv Syst Dis* 2014; 6: 41–49.

26. Polman CH, O'Connor PW, Havrdova E, et al. A randomized, placebo-controlled trial of natalizumab for relapsing multiple sclerosis. *N Engl J Med* 2006; 354: 899–910.

27. Gold R, Kappos L, Arnold DL, et al. Placebo-controlled phase 3 study of oral BG-12 for relapsing multiple sclerosis. *N Engl J Med* 2012; 367: 1098–1107.

28. Fox RJ, Miller DH, Phillips JT, et al. Placebo-controlled phase 3 study of oral BG-12 or glatiramer in multiple sclerosis. *N Engl J Med* 2012; 367: 1087–1097.

29. Vickers AJ, van Calster B, Steyerberg EW. A simple, step-by-step guide to interpreting decision curve analysis. *Diagn Progn Res* 2019; 3: 18.

30. Simpson EH. The Interpretation of Interaction in Contingency Tables. *J R Stat Soc Ser B Methodol* 1951; 13: 238–241.

31. Ameringer S, Serlin RC, Ward S. Simpson's Paradox and Experimental Research. *Nurs Res* 2009; 58: 123–127.

32. R Core Team. A language and environment for statistical computing. R Foundation for Statistical Computing, Vienna, Austria. URL https://www.R-project.org/.

33. Vickers AJ, Calster BV, Steyerberg EW. Net benefit approaches to the evaluation of prediction models, molecular markers, and diagnostic tests. *BMJ* 2016; 352: i6.

34. Kent DM, Rothwell PM, Ioannidis JP, et al. Assessing and reporting heterogeneity in treatment effects in clinical trials: a proposal. *Trials* 2010; 11: 85.

35. Vickers AJ, Cronin AM, Elkin EB, et al. Extensions to decision curve analysis, a novel method for evaluating diagnostic tests, prediction models and molecular markers. *BMC Med Inform Decis Mak* 2008; 8: 53.

36. Schmitz S, Maguire Á, Morris J, et al. The use of single armed observational data to closing the gap in otherwise disconnected evidence networks: a network meta-analysis in multiple myeloma. *BMC Med Res Methodol* 2018; 18: 66.

37. Thom HHZ, Capkun G, Cerulli A, et al. Network meta-analysis combining individual patient and aggregate data from a mixture of study designs with an application to pulmonary arterial hypertension. *BMC Med Res Methodol* 2015; 15: 34.

38. Signorovitch JE, Wu EQ, Yu AP, et al. Comparative effectiveness without head-to-head trials: a method for matching-adjusted indirect comparisons applied to psoriasis treatment with adalimumab or etanercept. *PharmacoEconomics* 2010; 28: 935–945.

39. Caro JJ, Ishak KJ. No head-to-head trial? simulate the missing arms. *PharmacoEconomics* 2010; 28: 957–967.

40. Lin L, Zhang J, Hodges JS, et al. Performing Arm-Based Network Meta-Analysis in R with the pcnetmeta Package. *J Stat Softw* 2017; 80: 5.

41. Donegan S, Williamson P, D'Alessandro U, et al. Assessing key assumptions of network meta-analysis: a review of methods. *Res Synth Methods* 2013; 4: 291–323.

42. Rouse B, Chaimani A, Li T. Network Meta-Analysis: An Introduction for Clinicians. *Intern Emerg Med* 2017; 12: 103–111.

43. Jansen JP, Naci H. Is network meta-analysis as valid as standard pairwise meta-analysis? It all depends on the distribution of effect modifiers. *BMC Med* 2013; 11: 159.

44. Caldwell DM. An overview of conducting systematic reviews with network meta-analysis. *Syst Rev* 2014; 3: 109.

# 7 Tables

**Table 1 Reaching the recommended treatment, via a prognostic model, between four options: placebo, glatiramer acetate, dimethyl fumarate, and natalizumab. Hypothetical Example in relapsing-remitting multiple sclerosis.**

| Treatment | Placebo | Glatiramer acetate | Dimethyl fumarate | Natalizumab |
|---|---|---|---|---|
| **Predicted risk to relapse within two years ($R_{i,j}$)** | 75% | 66% | 52% | 44% |
| **Predicted risk difference vs placebo ($RD_{i,j}$)** | - | 9% | 23% | 31% |
| **Threshold value for treatment $j$ ($T_j$)** | | 10% | 10% | 20% |
| $RD_{i,j} - T_j$ | | -1% | **12%** | 11% |
| **Recommended treatment via the prediction model** | | | Dimethyl fumarate | |

**Table 2 Detailed description of net benefit (NB) estimation for 'treat all with treatment j', and 'treat patients according to the prediction model' strategies on the RRMS example. The threshold values for glatiramer acetate, dimethyl fumarate, and natalizumab are noted as $T_{GA}, T_{DF}$, and $T_N$ respectively.**

|  | Treat all with | | | | Treat according to the model | | | | |
|---|---|---|---|---|---|---|---|---|---|
|  | Placebo ('treat none') | Glatiramer acetate | Dimethyl fumarate | Natalizumab | Placebo ('treat none') | Glatiramer acetate | Dimethyl fumarate | Natalizumab | Total |
| **Treatment rate** | 0% | 100% | 100% | 100% | $p_0^{Data_s}$ | $p_1^{Data_s}$ | $p_2^{Data_s}$ | $p_3^{Data_s}$ | $\sum_{j=0}^{J} p_j^{Data_s}$ |
| **Event rate** | $\hat{\varepsilon}_0$ as a meta-analysis of all placebo arms in $Data$ | $\hat{\varepsilon}_1 = \hat{\varepsilon}_0 \times RR_1^{Data}$ | $\hat{\varepsilon}_2 = \hat{\varepsilon}_0 \times RR_2^{Data}$ | $\hat{\varepsilon}_3 = \hat{\varepsilon}_0 \times RR_3^{Data}$ | $\hat{\varepsilon}_{s,0}$ as a meta-analysis of all placebo arms in $Data_s$ | $\hat{\varepsilon}_{s,1} = \hat{\varepsilon}_{s,0} \times RR_1^{Data_s}$ | $\hat{\varepsilon}_{s,2} = \hat{\varepsilon}_{s,0} \times RR_2^{Data_s}$ | $\hat{\varepsilon}_{s,3} = \hat{\varepsilon}_{s,0} \times RR_3^{Data_s}$ | $\hat{\varepsilon}_s = \sum_{j=0}^{J} p_j^{Data_s} \times \hat{\varepsilon}_{s,j}$ |
| **Decrease in event rate** | 0 | $\hat{\varepsilon}_0 - \hat{\varepsilon}_1$ | $\hat{\varepsilon}_0 - \hat{\varepsilon}_2$ | $\hat{\varepsilon}_0 - \hat{\varepsilon}_3$ | $\hat{\varepsilon}_0 - \sum_{j=0}^{J} p_j^{Data_s} \times \hat{\varepsilon}_{s,j}$ | | | | |
| **Net strategy Benefit** | 0 | $NB_1 = \hat{\varepsilon}_0 - \hat{\varepsilon}_1 - T_{GA}$ | $NB_2 = \hat{\varepsilon}_0 - \hat{\varepsilon}_2 - T_{DF}$ | $NB_3 = \hat{\varepsilon}_0 - \hat{\varepsilon}_3 - T_N$ | $NB_{model} = \hat{\varepsilon}_0 - \sum_{j=0}^{J} p_j^{Data_s} \times \hat{\varepsilon}_{s,j} - \sum_{j=1}^{J} p_j^{Data_s} \times T_j$ | | | | |

**Table 3 Net benefit (NB) estimation for each strategy on the multiple sclerosis example: 'treat none', 'treat all patients with glatiramer acetate', 'treat all patients with dimethyl fumarate', 'treat all patients with natalizumab', and 'treat patients according to the prediction model'. The threshold values used for the NB estimation are: $T_{GA} = T_{DF} = 10\%$ and $T_N = 20\%$ respectively.**

|  | Treat all with | | | | Treat according to the model | | | | |
| --- | --- | --- | --- | --- | --- | --- | --- | --- | --- |
|  | Placebo ('treat none') | Glatiramer acetate | Dimethyl fumarate | Natalizumab | Placebo ('treat none') | Glatiramer acetate | Dimethyl fumarate | Natalizumab | Total |
| **Treatment rate** | 0% | 100% | 100% | 100% | 4/652=0.6% | 0/652=0% | 418/652=64.1% | 230/652=35.3% | 100% |
| **RR from congruent dataset** | - | 0.68 | 0.59 | 0.52 | - | - | 0.24 | 0.40 |  |
| **Event rate** | $\hat{\varepsilon}_0 =$ 53% | $\hat{\varepsilon}_1 =$ 36% | $\hat{\varepsilon}_2 =$ 31% | $\hat{\varepsilon}_3 =$ 28% | $\hat{\varepsilon}_{s,0} =$ 75% | — | $\hat{\varepsilon}_{s,2} =$ 18% | $\hat{\varepsilon}_{s,3} =$ 30% | $\hat{\varepsilon}_s =$ 23% |
| **Decrease in event rate** | 0 | 17% | 22% | 25% |  |  | 30% |  |  |
| **Net strategy Benefit** | 0 | $NB_1 =$ 0.07 | $NB_2 =$ 0.12 | $NB_3 =$ 0.05 |  |  | $NB_{model} =$ 0.17 |  |  |

RR: Risk ratio

## 8 Figure legends

**Figure 1** Net-graph: Treatments compared in each one the available RCTs: AFFIRM, DEFINE, and CONFIRM.[26, 27, 28]

**Figure 2** Estimated probability to relapse within the next two years as a function of the baseline risk score. The x-axis shows the baseline risk score of relapsing within the next two years and the y-axis shows the estimated probability to relapse within the next two years under each one of the treatments. Between the two dashed vertical lines are the baseline risk values observed in the data used

**Figure 3** Decision curve analysis plot, for a range of threshold values for natalizumab (19%-40%) and equal constant threshold values for dimethyl fumarate and glatiramer acetate (10%). X-axis represents the range of threshold values for natalizumab, and y-axis represents the net benefit for each one of the five strategies: 'Treat none', 'treat based on the model', 'treat all with natalizumab', 'treat all with dimethyl fumarate' and 'treat all with glatiramer cetate'. (Dashed black line represents the highest net benefit)

**Figure 4** Heat plot for the decision curve analysis, in a range of threshold values. The same threshold is assumed for dimethyl fumarate (DF) and glatiramer acetate (GA) (4%-25%). The threshold values for natalizumab range between 19%-40%. The plot shows which approach has the highest net benefit between all possible approaches: a) Treat all patients with placebo, b) Treat all patients with natalizumab (N), c) Threat all patients with dimethyl fumarate, d) Treat all patients with glatiramer acetate, and e) Treat patients based on the prediction model. The empty grey cells present the threshold values combinations are not clinically possible. The numbers in the cells are differences in net benefit (NB) between the two strategies. When the 'treat patients based on the prediction model' strategy is the best, the number in the cell is the difference between its NB and the NB of the second-best strategy. Otherwise, we present the difference between the NB of the best strategy and the NB of the 'treat patients based on the prediction model' strategy. The presented NB estimations are multiplied with 100.

## 9 Figures

**Figure 1**

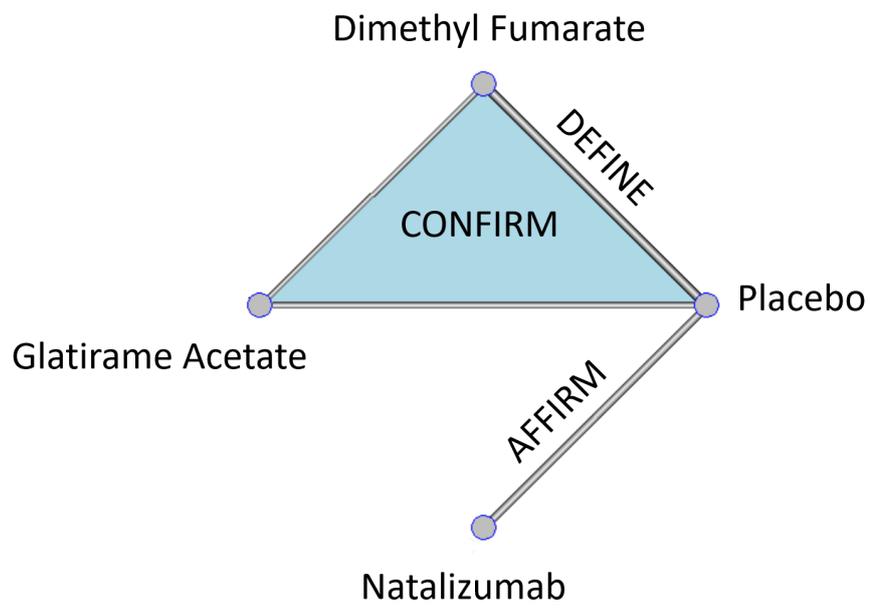

**Figure 2**

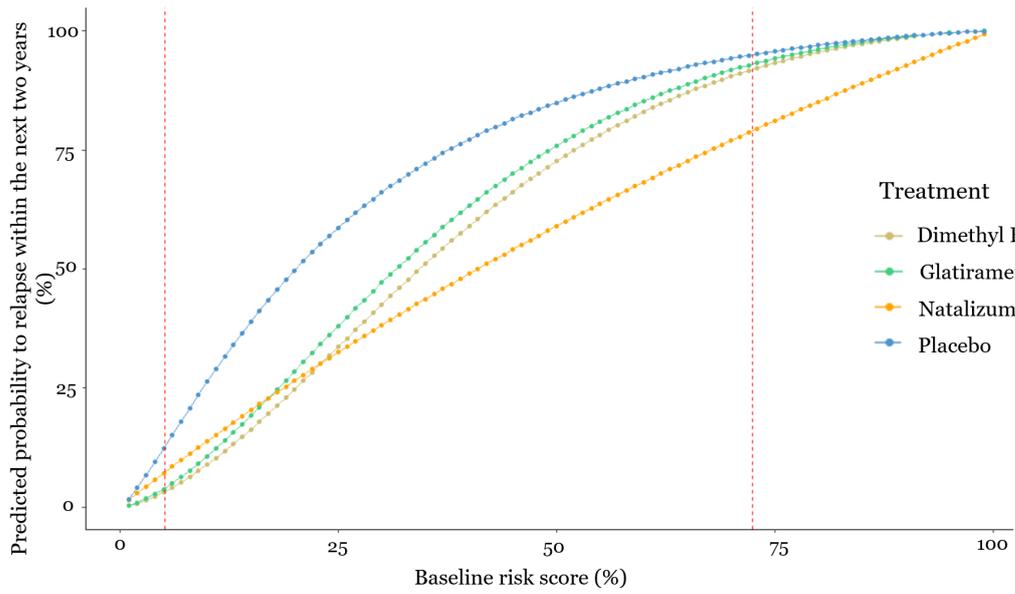

**Figure 3**

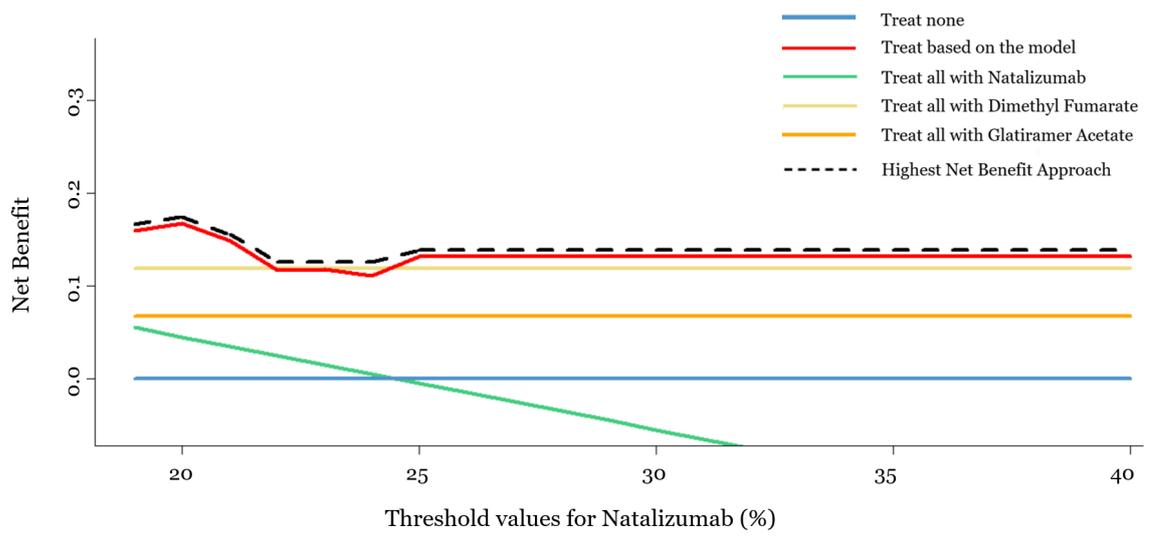

**Figure 4**

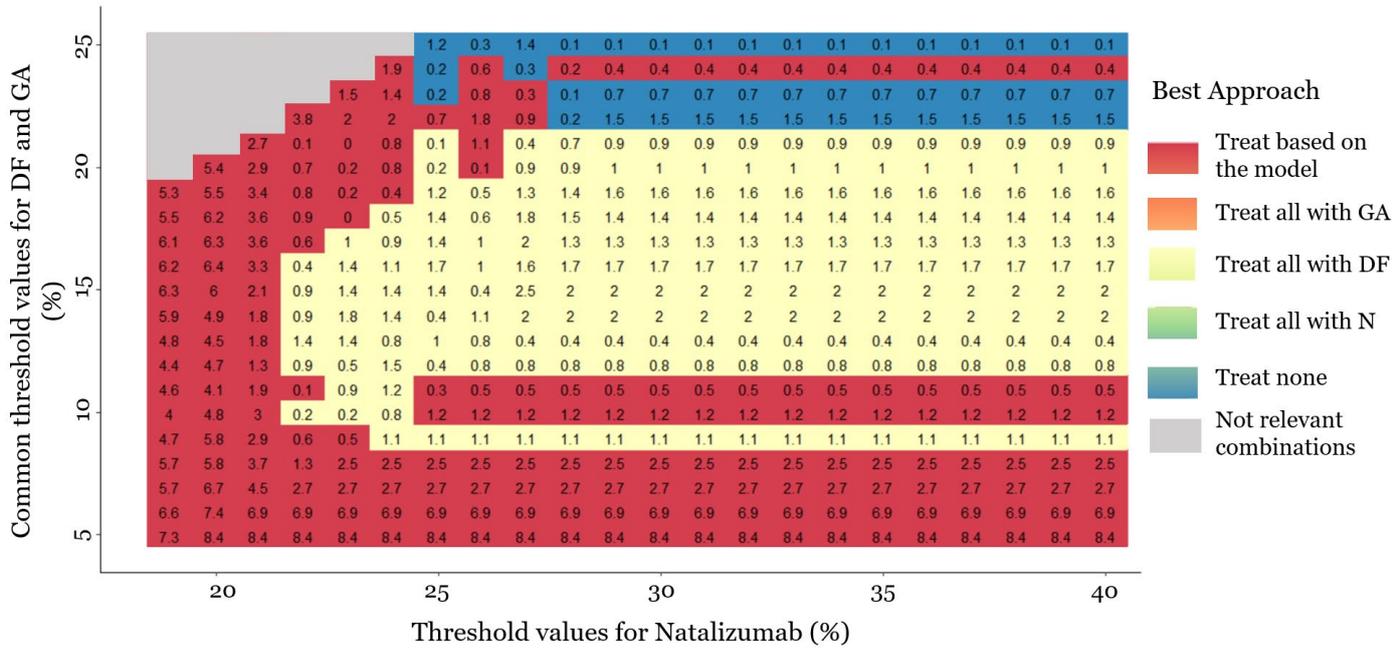

## 10 Appendix

Here, we describe in detail the fictional example, whose results are presented in **Table 3**. We describe in detail how to estimate the NB of each strategy for the set of threshold values $T_N = 20\%$, $T_{DF} = T_{GA} = 10\%$ (following all the steps presented in **Table 2**).

First, we describe the NB estimations for 'treat none', and 'treat all with treatment $j$' strategies, i.e., for 'treat all with glatiramer acetate', 'treat all with dimethyl fumarate, and 'treat all with natalizumab' strategies (first 4 columns in **Table 2**). We need to estimate the treatment rates for each strategy. For strategy 'treat none', the treatment rate is 0% as none of the individuals will take any treatment. For strategy 'treat all with glatiramer acetate', the treatment rate will be 100% for glatiramer acetate (as all individuals will take this treatment) and 0% for any other treatment. Similarly, the treatment rate for dimethyl fumarate will be 100% for 'treat all with dimethyl fumarate' and 0% for other treatment options, and under 'treat all with natalizumab' strategy natalizumab's treatment rate will be 100% and 0% for any other treatment option. Then, we estimate each treatment's risk ratio versus placebo for each strategy, via NMA. Here, for 'treat all with treatment $j$' strategies, we use the entire dataset (i.e., 2000 patients) and we perform NMA to estimate the corresponding risk ratios to relapse. The risk ratio of glatiramer acetate versus placebo is 0.68, the risk ratio of dimethyl fumarate vs placebo is 0.59, and the risk ratio of natalizumab vs placebo is 0.52. Then, the estimation of the event rates and the decrease in event rates compared to placebo ('treat none') follows. We estimate the event rate (i.e., the rate of patients that relapsed) under placebo as a meta-analysis of all placebo arms in the entire dataset, which is equal to 53%. Then, the event rate under glatiramer acetate can be estimated as $\hat{\varepsilon}_1 = \hat{\varepsilon}_0 \times RR_1^{Data} = 53\% \times 0.68 \approx 36\%$. Hence, if we treat all with glatiramer acetate we expect a decrease in event rate (compared to treat none) equal to $\hat{\varepsilon}_0 - \hat{\varepsilon}_1 = 53\% - 36\% = 17\%$. This means that we expect 17 patients per 100 less to relapse if we use the strategy 'treat all with glatiramer acetate' compared to use the strategy 'treat none'. Similarly, the event rate under dimethyl fumarate is $\hat{\varepsilon}_2 = \hat{\varepsilon}_0 \times RR_2^{Data} = 53\% \times 0.59 \approx 31\%$ and the expected decrease in the event rate if we treat all patients with dimethyl fumarate is $\hat{\varepsilon}_0 - \hat{\varepsilon}_2 = 53\% - 31\% = 22\%$. Finally, the event rate under natalizumab is $\hat{\varepsilon}_3 = \hat{\varepsilon}_0 \times RR_3^{Data} = 53\% \times 0.52 \approx 28\%$ and the expected decrease in the event rate if we treat all patients with natalizumab is $\hat{\varepsilon}_0 - \hat{\varepsilon}_3 = 53\% - 28\% = 25\%$. Now, we can estimate the NBs for each 'treat all with treatment $j$' strategy. Let us assume a threshold value for natalizumab equal to $T_N = 20\%$, meaning that we assume a relapse 5 times worse than the potential side effects of natalizumab, and for dimethyl fumarate and glatiramer acetate we set threshold values equal to $T_{DF} = T_{GA} = 10\%$, meaning we assume a relapse 10 times worse than the potential side effects of these treatments. Then, the NB (which outweighs the benefits, i.e., decrease in event rate $\hat{\varepsilon}_0 - \hat{\varepsilon}_s$, and the risks of each strategy, i.e., the percentage of patients taking each treatment multiplied with the corresponding threshold value) for 'treat all with glatiramer acetate' strategy will be $NB_1 = \hat{\varepsilon}_0 - \hat{\varepsilon}_1 - (100\%)T_{GA} = 17\% - (100\%) \times 10\% = 0.07$. Similarly, for 'treat all with dimethyl fumarate' strategy: $NB_2 = \hat{\varepsilon}_0 - \hat{\varepsilon}_2 - (100\%)T_{DF} = 22\% - (100\%) \times 10\% = 0.12$, and for strategy 'treat all with natalizumab' will be $NB_3 = \hat{\varepsilon}_0 - \hat{\varepsilon}_3 - (100\%)T_N = 25\% - (100\%) \times 20\% = 0.05$.

For 'treat patients according to the prediction model' strategy, first the model need to make treatment recommendations for each patients (section 3.2). The prediction model will recommend a treatment to each patient based on the threshold values set. For example, if- $T_N = 20\%$ (meaning that a patient would not take natalizumab if its predicted risk difference is less than 20% under natalizumab compared to placebo), and $T_{DF} = T_{GA} = 10\%$, will recommend any of these options only if the predicted decrease in the probability to relapse compared to placebo exceeds the corresponding threshold value. For example, let's think about a patient, who under natalizumab has a predicted probability to relapse equal to 25%, whereas under placebo the patient has 37% predicted probability to relapse. The expected risk difference under natalizumab for this specific patient is 37%-25%=12%, that is less than natalizumab's threshold value, and the model would not recommend this treatment to this patient. Let's also assume that the same patient has a predicted risk difference equal to 17% under dimethyl fumarate and 14% under glatiramer acetate. Both options give higher predicted risk difference than the corresponding threshold values. Between these two treatment options, the model will recommend the treatment with the maximum difference between the predicted risk difference and the corresponding threshold value, i.e., $max\{(17\% - 10\%), (14\% - 10\%)\} = max\{7\%, 4\%\} = 7\%$. Hence, in this case, the model would recommend dimethyl fumarate to this patient. In case for a patient all the predicted risk differences are lower than the corresponding threshold values, the model would recommend placebo.

After defining via the model which treatment will be recommended to each patient, we can estimate all the required quantities for the NB estimation 'treat patients according to the prediction model' strategy (last 5 columns in Table 2). First, we need the congruent dataset, which includes the patients whose the recommended treatment via the prediction model is the same as the actual given one. The congruent dataset includes 652 patients; 4 patients in placebo, 418 in dimethyl fumarate, and 230 in natalizumab. Therefore, the corresponding treatment rates under 'treat patients according to the prediction model' strategy will be 4/652=0.6% for placebo, 418/652=64.1% for dimethyl fumarate, and 230/652=35.3% for natalizumab. Then, we perform NMA in the congruent dataset (e.g., 652 patients) to estimate each treatment's risk ratio versus placebo. The estimated risk ratio for dimethyl fumarate vs placebo is 0.24 and for natalizumab vs placebo is 0.40. Note that as the model does not recommend glatiramer acetate to any patient, the risk ratio for this treatment is not estimated. Then, we estimate the event rate under placebo as a meta-analysis of all placebo arms in the congruent dataset, which is equal to 75%. The event rate under dimethyl fumarate is $\hat{\varepsilon}_{s,1} = \hat{\varepsilon}_{s,0} \times RR_1^{Data_s} = 75\% \times 0.24 = 18\%$, and the event rate under natalizumab is $\hat{\varepsilon}_{s,2} = \hat{\varepsilon}_{s,0} \times RR_2^{Data_s} = 75\% \times 0.40 = 30\%$. Then we need to estimate the total event rate under 'treat patients according to the prediction model' strategy that is $\hat{\varepsilon}_s = \sum_{j=0}^{J} p_j^{Data_s} \times \hat{\varepsilon}_{s,j} = 64.1\% \times 18\% + 35.3\% \times 30\% \approx 23\%$. Then, the estimated decrease in event rate if we use the 'treat patients according to the prediction model' strategy compared to 'treat none' strategy is $\hat{\varepsilon}_0 - \hat{\varepsilon}_s = 53\% - 23\% = 30\%$, meaning that we expect 30 patients per 100 less to relapse if we use this strategy 'compared to use the strategy 'treat none'. Finally, the NB for 'treat patients according to the prediction model' strategy will be $NB_s = \hat{\varepsilon}_0 - \hat{\varepsilon}_s - \sum_{j=1}^{J} p_j^{Data_s} \times T_j = 30\% - (64.1\% \times 10\% + 35.3\% \times 20\%) \approx 30\% - 13\% \approx 0.17$.

Finally, we compare the estimated NBs of each strategy, and the strategy with the highest NB is the most clinical useful one strategy. Here, the highest NB is this of the 'treat patients according to the prediction model' strategy, with NB=0.17, meaning that the 'treat patients according to the prediction model' strategy would lead to 17% fewer patients that relapsed compared to 'treat none' strategy.

Note that, the model's clinical utility needs to be validated into a reasonable range of threshold values (not only for one set of threshold values). Therefore, this procedure needs to be repeated for all the threshold values into the defined threshold values range.